\documentclass{webofc}

\usepackage[varg]{txfonts}   
\hypersetup{colorlinks=true,citecolor=blue,urlcolor=blue,linkcolor=blue}
\usepackage{tikz}
\usetikzlibrary{shapes,arrows,positioning,fit,backgrounds,calc}
\usepackage{caption}
\usepackage{subcaption}
\usepackage{float}
\usepackage{wrapfig}

\usepackage{mathtools}
\usepackage{gensymb}
\usepackage{slashed}

\usepackage{pgffor}   
\usepackage{ifthen}   
\usepackage{xcolor}

\begin{document}
\title{Measurement of beam-polarized Deeply Virtual Compton Scattering observables with the CLAS12 detector at Jefferson Lab}
\author{\firstname{Juan Sebastian} \lastname{Alvarado}\inst{1}\fnsep\thanks{\email{jsalvaradog@outlook.com}} \and
        \firstname{Mostafa} \lastname{Hoballah}\inst{1}\fnsep\and
        \firstname{Eric} \lastname{Voutier}\inst{1}\fnsep
}
\institute{Université Paris-Saclay, CNRS/IN2P3, IJCLab, 91405 Orsay, France}
\abstract{In the context of nucleon structure studies, Generalized Parton Distributions (GPDs) are crucial for understanding the correlation between the longitudinal momentum and the transverse position of partons inside the nucleon. A privileged channel for GPD studies is the Deeply Virtual Compton Scattering (DVCS) process, whose experimental observables can provide access to GPDs through spin-dependent asymmetries. Although detecting all final state particles is preferred for selecting DVCS events, DVCS identification can be ensured by requiring the detection of only two final state particles, as the missing particle can be reconstructed from conservation laws. In this work, we present preliminary Beam Spin Asymmetry and cross-section measurements of proton-DVCS in the $e\gamma$ topology from experimental data taken by the CLAS12 detector at Jefferson Lab. Moreover, we show that relying on $e\gamma$ detection and Machine Learning techniques boosts statistics and gives access to a wider phase space than the proton-detected topology, providing the first time CLAS12 measurement of this observable in the $-t$ range below $0.1$ GeV$^{2}$.
}
\maketitle
\section{Introduction}
\begin{figure}[H]
    \centering
    \begin{subfigure}[b]{0.32\textwidth}
        \centering
        \includegraphics[width=0.75\textwidth]{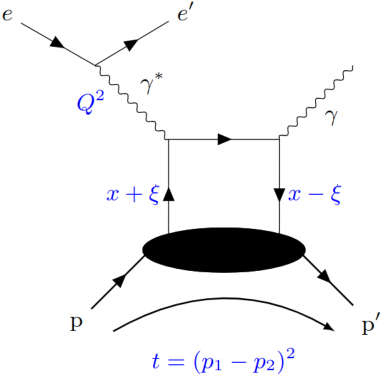}
        \caption{DVCS handbag diagram.}
    \end{subfigure}
    \begin{subfigure}[b]{0.32\textwidth}
        \centering
        \includegraphics[width=\textwidth]{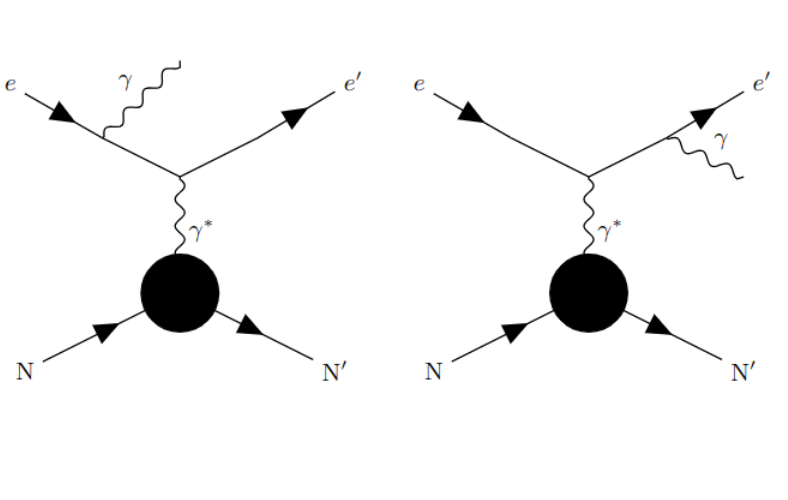}
        \caption{Bethe Heitler process.}
    \end{subfigure}
    \begin{subfigure}[b]{0.32\textwidth}
        \centering
        \includegraphics[width=\textwidth]{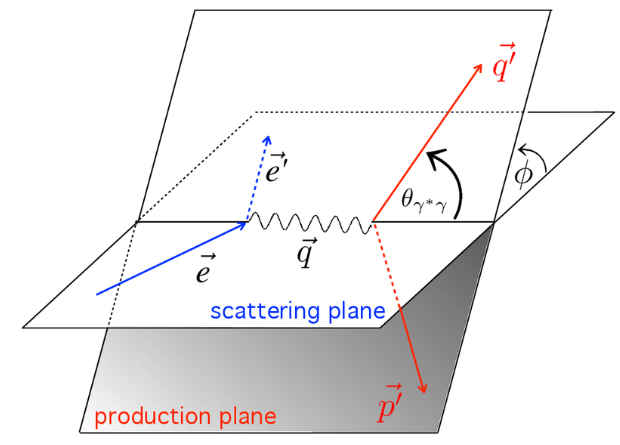}
        \caption{DVCS kinematics planes.}
        \label{Planes}
    \end{subfigure}
    \caption{Processes entering the $ep\rightarrow ep\gamma$ reaction. $Q^{2}$ denotes the photon virtuality and $\phi$ the angle between the scattering and production planes.}
    \label{DVCS}
\end{figure}

In the context of nucleon structure studies, Generalized Parton Distributions (GPDs) provide a unified framework to Form Factors and Parton Distribution Functions, encoding nucleon properties arising from the correlation between the longitudinal momentum and transverse spatial position of partons \cite{GPD4}. Among other properties, they enable nucleon tomography \cite{GPD3} and access the contribution of the orbital momentum of quarks to the nucleon total spin through Ji's sum rule \cite{GPD5}. Deeply Virtual Compton Scattering (DVCS) is widely regarded as the golden channel for chiral-even GPD studies, as its beam and target spin polarized observables have a direct interpretation in terms of GPDs. GPDs enter the cross section of the $ep\rightarrow e\gamma p$ reaction in the form of Compton Form Factors (CFFs) given by:
\begin{align}
    \mathcal{F}(\xi , t) &=\mathcal{P}\int_{-1}^1dx'~F(x',\xi,t)\bigg[\frac{1}{x'-\xi}\pm\frac{1}{x'+\xi}\bigg] -i\pi\big[F(\xi,\xi,t)-F(-\xi,\xi,t)\big], \label{CFF}
\end{align}
\noindent
where $\mathcal{P}$ denotes Cauchy's principal value, $F=H,E,\widetilde{H}, \widetilde{E}$ stands for the four chiral-even GPDs of a spin $1/2$  particle, $\xi$ is the skewness parameter, $t$ is the momentum transfer to the nucleon. The top and bottom signs apply to the unpolarized ($H$, $E$) and polarized ($\widetilde{H}$, $\widetilde{E}$) GPDs respectively. Access to GPDs is granted by observables sensitive to the imaginary part of CFFs, such as the Beam Spin Asymmetry (BSA) $A_{LU}(\phi)$ arising from the interference of the DVCS and Bethe-Heitler mechanisms (Fig. \ref{DVCS}).
The CLAS12 detector \cite{CLAS12}, located in the Hall B of the Thomas Jefferson National Accelerator Facility (JLab), is a large acceptance spectrometer designed to detect, among other particles, electrons, photons, and protons across a wide kinematic range. This DVCS analysis relies on the data collected by the CLAS12 detector during its first commissioning period in fall 2018. The experimental setup consisted of a polarized electron beam impinging on an unpolarized liquid hydrogen target \cite{Maxime}.


\section{Channel selection}
The selection of DVCS candidates is based on kinematic thresholds of $1$ GeV/$c$ for electrons, $2$ GeV/$c$ for photons and, when detected, $0.3$ GeV/$c$ for protons. All events are further constrained to the deep inelastic scattering regime by requiring $W > 2$ GeV and $Q^{2} > 1$ GeV$^{2}$, where $W$ denotes the center-of-mass energy. Let $\Delta t$ represent the difference in the nucleon momentum transfer $t$ calculated via two independent methods, while $\Delta\phi$ denotes the difference between the azimuthal angles of the lepton and hadron planes (Fig. \ref{Planes}). DVCS candidates are selected by requiring $|\Delta t| < 2$ GeV$^{2}$, $|\Delta\phi| < 2\degree$, and a missing momentum of the $ep \rightarrow e\gamma p$ reaction $P_{\text{miss}} < 1$ GeV/$c$.
 \begin{figure}[H]
    \centering
    \includegraphics[width=0.6\textwidth]{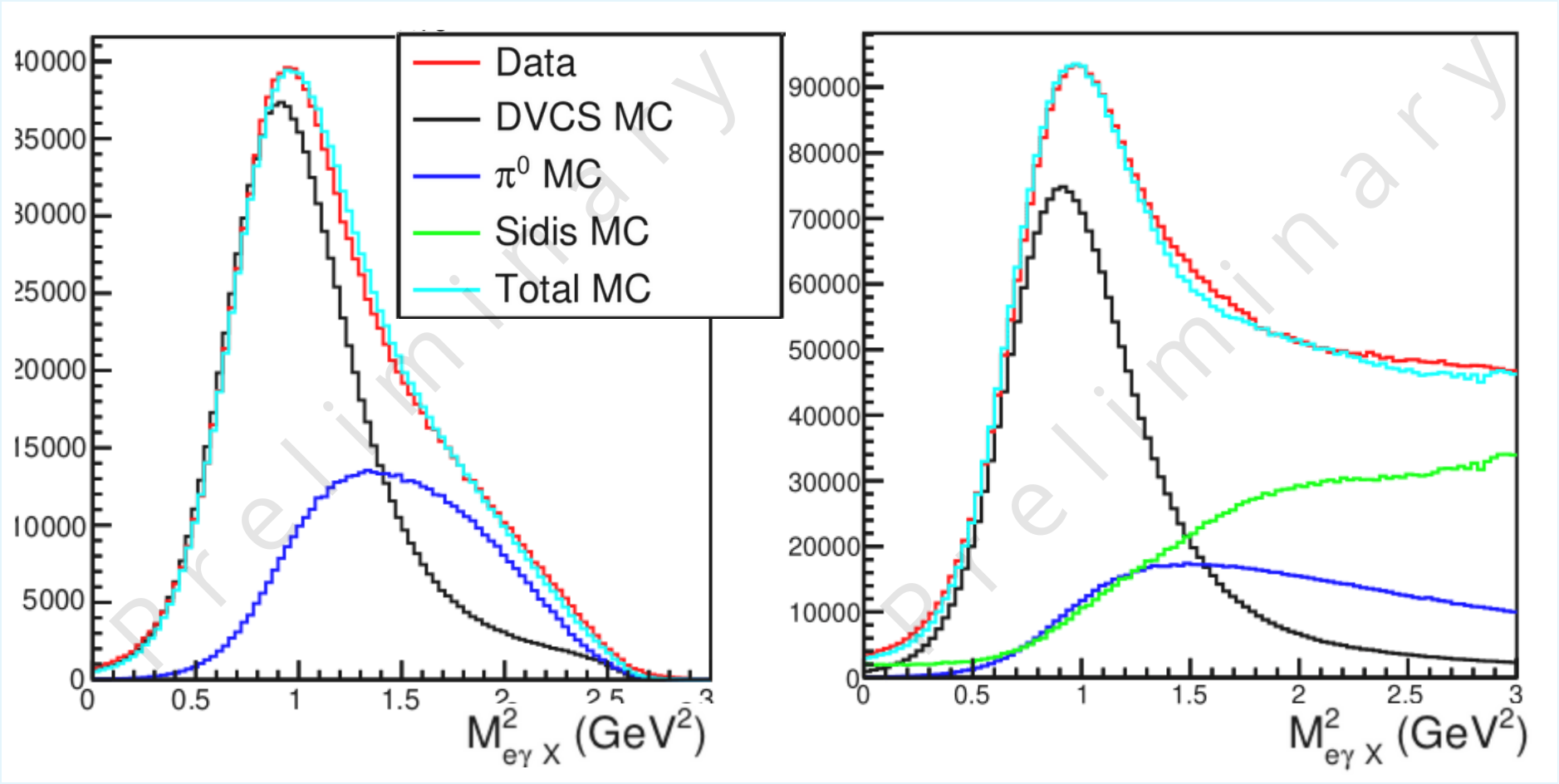}
    \caption{Missing mass of the $ep\rightarrow e\gamma X$ reaction when the detected proton information is used (left) or not (right).}
    \label{Selection}
\end{figure}

The resulting missing proton mass spectrum, shown in Fig. \ref{Selection}, suggests that the dominant background to the $e\gamma p$ final state arises from exclusive $\pi^{0}$ production, while the $e\gamma$ topology additionally includes a contribution from semi-inclusive DIS (SIDIS). In both cases, the contamination originates from a partially reconstructed $\pi^{0}\rightarrow\gamma(\gamma)$  decay where one photon is energetic enough to mimic a DVCS event and passes the selection criteria.  Background contributions are further suppressed with a Boosted Decision Tree (BDT) classifier trained with simulated samples of DVCS (signal) and exclusive $\pi^{0}$ production (background). Residual contamination after BDT classification is subtracted with standard background subtraction techniques developed by the CLAS12 collaboration.

\section{BSA measurements}
\begin{figure}[H]
    \centering
    \includegraphics[width=\linewidth]{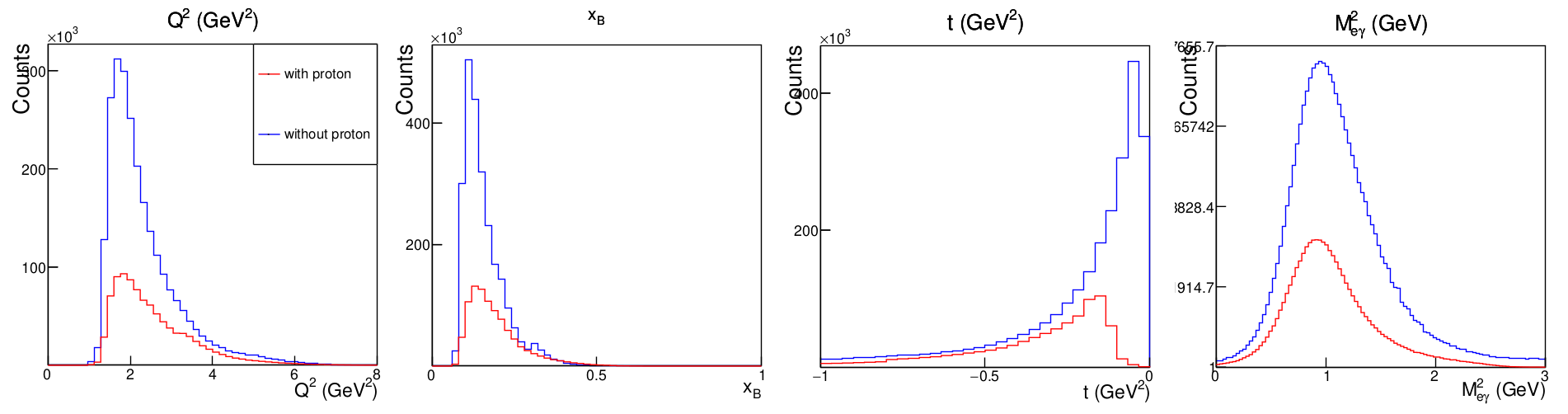}
    \caption{Final $Q^{2}$, $x_{B}$, $t$ and missing proton mass distributions. $x_{B}$ denotes the Bjorken scaling variable.}
    \label{Result_NP}
\end{figure}

A direct comparison of the $t$, $Q^{2}$ and $x_{B}$ distributions for the $e\gamma p$ and $e\gamma$ topologies (Fig. \ref{Result_NP}) demonstrates that lifting the requirement of proton detection provides a significant increase of statistics across the full kinematic range. In particular, the $e\gamma$ topology provides a CLAS12 first-time access to the small $t$ region, which remains inaccessible for an $e\gamma p$ analysis due to the limited  CLAS12 acceptance for low-momentum recoil protons.


\begin{figure}[H]
    \centering
    \includegraphics[width=\textwidth]{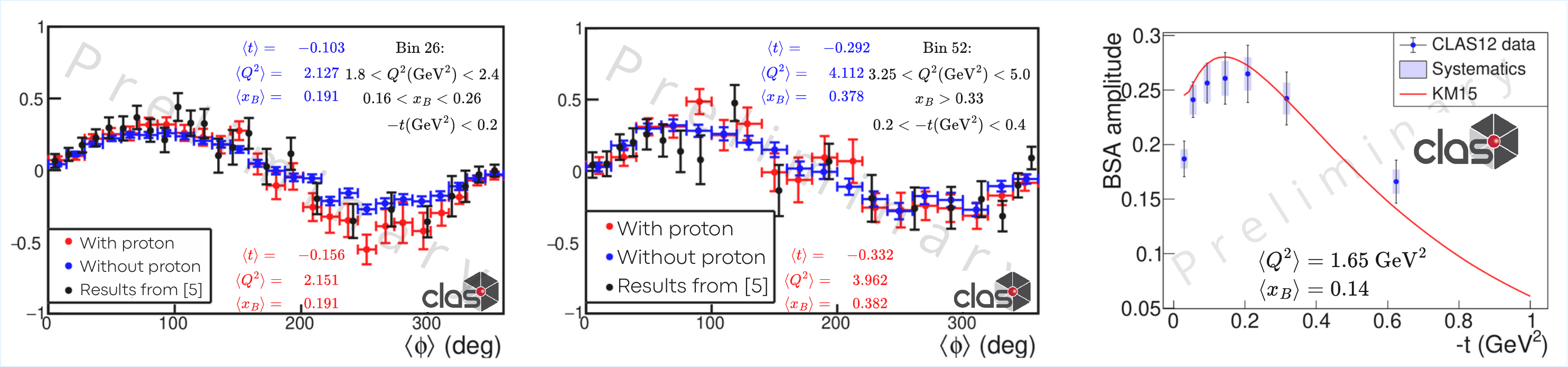}
    \caption{Sample BSA measurements compared to the published results \cite{Maxime} (left and middle plots). BSA amplitude as a function of $t$  at $\langle Q^{2}\rangle=1.65$ GeV$^{2}$ and $\langle x_{B} \rangle=0.14$ (right).}
    \label{BSA}
\end{figure}

Beam Spin Asymmetry (BSA) measurements are extracted as a function of $\phi$ (see Fig. \ref{Planes}) in bins of $Q^{2}$, $x_{B}$ and $t$, allowing for a multidimensional exploration of the DVCS phase space to access the imaginary part of the CFFs, and thus to GPDs. Provided a $P_{b}=85.92\pm 1.29~\%$ beam polarization in the experimental configuration, the BSA is defined as the normalized difference in yield for opposite beam helicities:
\begin{align}
    A_{LU}&=\frac{1}{P_{b}}\frac{N^{+} - N^{-}}{N^{+} + N^{-}}\propto \mathfrak{I} \mathfrak{m}\left\{F_{1} \mathcal{H}+\xi\left(F_{1}+F_{2}\right) \tilde{\mathcal{H}}-\frac{t}{4 M^{2}} F_{2} \mathcal{E}\right\},
\end{align}
\noindent
where $N^{\pm}$ represents the number of reconstructed DVCS events for positive and negative beam helicities, and $F_{1}$ and $F_{2}$ denote the nucleon form factors. Fig. \ref{BSA} presents a couple of sample preliminary comparisons of the $e\gamma$ and $e\gamma p$ results along with previously published results from the collaboration \cite{Maxime}, showing the agreement among all three results. Moreover, a representative plot of the BSA $\sin\phi$ moment as a function of $t$ is shown, illustrating the extent of the $e\gamma$ topology coverage in $t$ and the compatibility of the measurements with the KM15 model predictions \cite{KM15}.

\begin{figure}[H]
    \centering
    \begin{subfigure}[b]{0.44\textwidth}
        \centering
        \includegraphics[width=0.8\textwidth]{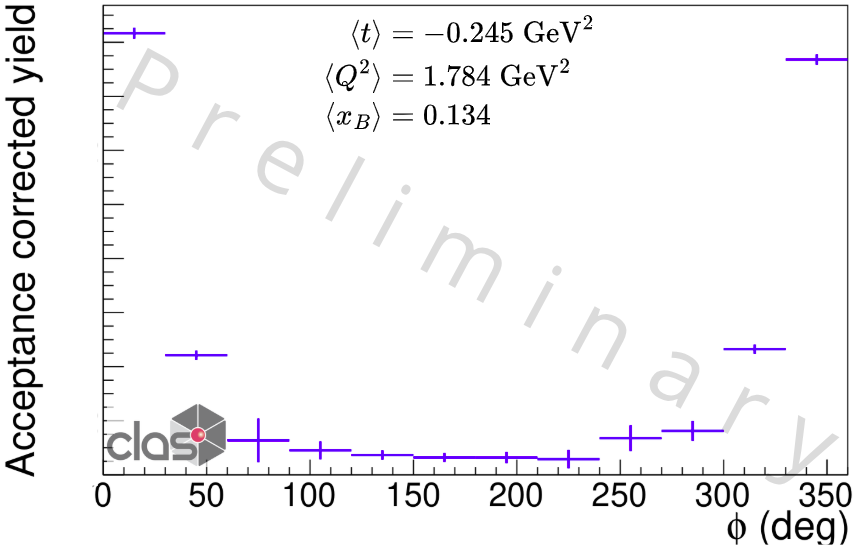}
    \end{subfigure}
    \begin{subfigure}[b]{0.44\textwidth}
        \centering
        \includegraphics[width=0.8\textwidth]{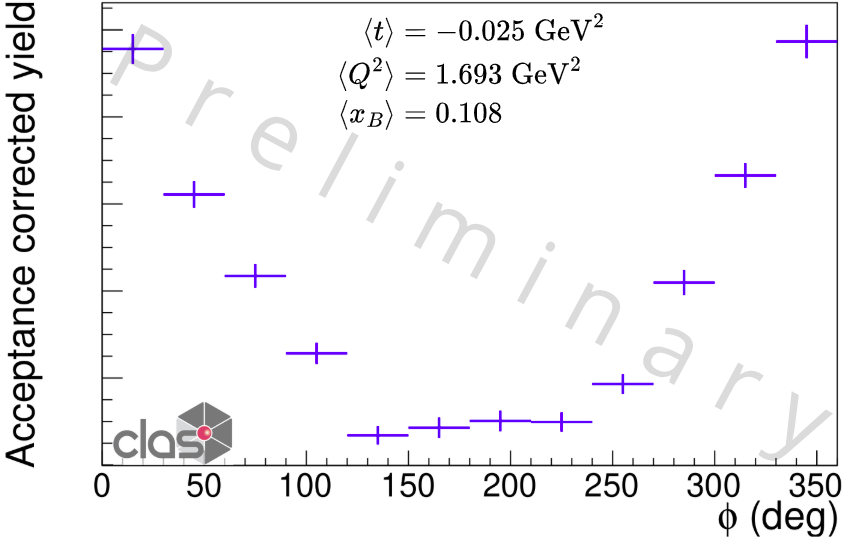}
    \end{subfigure}
    \caption{Preliminary acceptance corrected yields without requiring the recoil proton information.}
    \label{xs}
\end{figure}

In brief, these preliminary results demonstrate the feasibility and effectiveness of the $e\gamma$ detection to select DVCS events and extract BSA measurements at CLAS12. In particular, this semi-exclusive approach relying on BDT classification significantly improves statistical precision and access to the otherwise inaccessible low $-t$ region at CLAS12. The observed agreement between the two selection approaches and previously published results confirms the reliability of this method. Beyond BSA measurements, the achieved control over the background sources for this selection strategy lays the groundwork for future cross section extractions, further extending the physics reach of the CLAS12 DVCS program. As a representative example, Fig. \ref{xs} shows sample acceptance-corrected yield distributions exhibiting the expected behavior of the DVCS cross section. 

\noindent
\textbf{Acknowledgements}: This work was supported by the Programme blanc of the physics Graduate School of the PHENIICS doctoral school [Contract No. D22-ET11] and the french Centre National de la Recherche Scientifique (CNRS).


\begin{thebibliography}{99}
\bibitem{GPD4} M. V. Polyakov, Phys. Lett. B 555 (2003) 57.
\bibitem{GPD3} M. Burkardt, Phys. Rev. D 62 (2000) 071503.
\bibitem{GPD5}  X. Ji, Phys. Rev. Lett. 78 (1997) 610.
\bibitem{CLAS12} V. D. Burkert, Nucl. Instrum. Meth. A, 959 (2020) 163419
\bibitem{Maxime} G. Christiaens \textit{et al.}, Phys. Rev. Lett. 130 (2023) 211902.
\bibitem{KM15} K. Kumeri{\v{c}}ki and D. Müller, Nucl. Phys. B 841 (2010) 1.
 

\end{thebibliography}
\end{document}